\documentclass[reprint,
 amsmath,amssymb,
 aps,
 pra,showkeys,
]{revtex4-2}

\usepackage[dvipsnames]{xcolor}
\usepackage[colorlinks=true, linkcolor=blue, citecolor=blue, urlcolor=blue]{hyperref}
\usepackage[nameinlink]{cleveref}
\usepackage{lipsum}
\usepackage{amsthm}
\theoremstyle{remark} 
\newtheorem{remark}{Remark}
\usepackage{graphicx}
\usepackage{dcolumn}
\usepackage{bm}
\Crefname{section}{Section}{Sections}
\Crefname{appendix}{Appendix}{Appendices}
\usepackage{subcaption}
\usepackage{pgfplots}

\pgfplotsset{compat=1.18}

\begin{document}

\title{Effective Complexity Reduction of the Landau-de Gennes Elastic Energy: A Quantitative Framework and Numerical Validation}

% Old title
%\title{Numerical validation of scale separation regimes for the elastic constants\\in the Landau-de Gennes model for nematic liquid crystals}% Force line breaks with \\

% Other titles
% \title{Quantitative Criteria for Elastic Energy Reduction in Nematic Liquid Crystals: Bridging Scaling Theory and Computation}

% \title{A Practical Framework for Reducing the Multi-Constant Landau-de Gennes Model: Scaling Predictions and openQmin Validation}

% \title{Rigorous Thresholds for the One-Constant Approximation in the Landau-de Gennes Model}

\author{Razvan-Dumitru Ceuca}
%\email{razvan-dumitru.ceuca@academic.tuiasi.ro}
\affiliation{%
``Gh. Asachi" Technical University of Iasi, Bd. Carol I, nr. 11, 700506, Ia\c si, Romania
}%

\author{Simone Rusconi}
\affiliation{
 CUNEF Universidad, C/ de los Pirineos 55, 28040 Madrid, Spain
}%

\author{Arghir-Dani Zarnescu}
 %\altaffiliation[Also at ]{IMAR, Bucharest, Romania}%Lines break automatically or can be forced with \\
 \affiliation{%
 BCAM, Bilbao, Spain, and IMAR, Bucharest, Romania
}%

\begin{abstract}
Abstract: We revisit the elastic energy formulation of the Landau–de Gennes model for nematic liquid crystals, focusing on quantitative reductions of the multi-constant elastic energy. Building on the generalized optimal scaling procedure (GOS) introduced by Rusconi et al. in 2025, we identify explicit parameter regimes in which the three-constant model $(L_1,L_2,L_3)$ can be reduced to $(L_1,L_2,0)$ and how the two-constant model $(L_1,L_2,0)$ can be reduced to the commonly used one-constant configuration $(L_1,0,0)$. The analytical scaling predictions are tested numerically using the openQmin simulation framework, confirming that below a critical threshold for $L_3$ or $L_2$, given by GOS, the deviation from the reduced model remains of the same order of magnitude as predicted by the scaling theory. These results provide a quantitative criterion for the validity of reduced elastic models and establish a direct connection between optimal scaling arguments and numerical observations within the Landau–de Gennes framework.
\end{abstract}

%Use showkeys class option if keyword
\maketitle

%\tableofcontents

\section{Introduction}

Nematic liquid crystals exhibit a rich variety of topological defects and structural behaviors that are fundamentally governed by the delicate interplay between bulk elasticity, surface anchoring, and spatial confinement.

Within the continuum framework, the Landau-de Gennes (LdG) tensorial order parameter is widely utilized to capture both the macroscopic orientation and the degree of order, proving particularly crucial near defect cores where the standard director field description breaks down.

The elastic energy landscape in the full LdG model is governed by multiple elastic constants - typically denoted as $L_1$, $L_2$, and $L_3$ - which correspond to the splay, twist, and bend deformations inherent to the physical material. 

Simulating the full multi-constant elastic energy is both computationally expensive and mathematically highly complex. Consequently, we frequently reduce the multi-constant model to simplified configurations, with the most ubiquitous being the one-constant approximation, $(L_1,0,0)$. This reduction operates on the assumption that a single elastic constant is sufficient to capture the relevant physical phenomena, especially when bulk properties or surface interactions dominate the system's overall behavior. However, this sweeping simplification is overwhelmingly adopted as a pragmatic or ad-hoc computational choice, lacking a rigorous mathematical criterion to explicitly justify exactly when the neglected constants are physically inconsequential. 

This paper bridges the gap between abstract mathe\-matical scaling and practical numerical simulation by providing an actionable, quantitative framework for reducing the multi-constant elastic energy. Rather than simply proving that a simplified model can exist in the asymptotic limit, we provide an effective methodology detailing exactly how and when to safely simplify the elastic free energy. Building on the \textit{Generalized Optimal Scaling} (GOS) procedure \cite{rusconi25}, we identify explicit, parameter-dependent regimes that dictate the following:
\begin{itemize}
    \item The precise conditions under which the three-constant model $(L_1,L_2,L_3)$ can be safely reduced to $(L_1,L_2,0)$. %[cite: 2]
    \item The specific thresholds allowing the two-constant model $(L_1,L_2,0)$ to be further reduced to the one-constant configuration $(L_1,0,0)$. %[cite: 2]
\end{itemize}

By translating complex algebraic bounds into direct, physically observable thresholds, we establish a precise and rigorous rule for the validity of the one-constant approximation. %[cite: 2]

To thoroughly validate these analytical scaling predictions, we utilize the \texttt{open-Qmin} simulation framework to model a spherical colloidal particle confined within a nematic liquid crystal \cite{sussman19}. By systematically comparing the equilibrium states of the full and reduced models, we demonstrate that as long as the neglected elastic constant ($L_3$ or $L_2$) remains below the critical threshold established by the GOS method, the deviation between the simulated models remains strictly bounded and of the exact same order of magnitude as predicted by the scaling theory. Ultimately, these results provide computational physicists with a rigorous, ready-to-use diagnostic tool to confidently apply reduced elastic models without introducing unphysical structural artifacts into their simulations. 

\section{Framework used}

We consider a cubic container $[0,L]^3$ in which we place a spherical colloidal particle $B(C_0,R)$ in the center of the cube, with $L>0$, $R\in(0,L)$ and $C_0(L/2,L/2,L/2)$. We denote $\Omega=\big[0,L\big]^3\setminus B(C_0,R)$. Let also
\begin{align*}
\mathcal{S}_0=\{Q\in\mathbb{R}^{3\times 3}\;|\;Q=Q^T,\;\text{tr}(Q)=0\}
\end{align*}
the set of all $Q$-tensors. We denote by $H^1_{\#}(\Omega,\mathcal{S}_0)$ the $H^1$ Sobolev space of all maps $Q$ defined on $\Omega$ with values in $\mathcal{S}_0$ such that they are periodic on the cube $[0,L]^3$.

We consider the following free energy functional of the Landau-de Gennes type $\mathcal{F}:H^1(\Omega;\mathcal{S}_0)\to\mathbb{R}$:
\begin{align}\label{eq:min_problem}
\mathcal{F}(Q):=\int\limits_{\Omega}f_e(Q,\nabla Q)\;\text{d}x+\int\limits_{\Omega}f_b(Q)\;\text{d}x+\int\limits_{\partial B}f_s(Q,\nu)\;\text{d}\sigma,
\end{align}
where the elastic energy is given by:
\begin{align}\label{eq:elastic_energy_LdG}
f_e(Q,\nabla Q)&=\dfrac{L_1}{2}\dfrac{\partial Q_{ij}}{\partial x_k}\dfrac{\partial Q_{ij}}{\partial x_k}+\dfrac{L_2}{2}\dfrac{\partial Q_{ij}}{\partial x_j}\dfrac{\partial Q_{ik}}{\partial x_k}+\notag\\
&+\dfrac{L_3}{2}\dfrac{\partial Q_{ik}}{\partial x_j}\dfrac{\partial Q_{ij}}{\partial x_k}+\dfrac{L_4}{2}\epsilon_{lik}Q_{lj}\dfrac{\partial Q_{ij}}{\partial x_k}+\\
&+\dfrac{L_6}{2}Q_{lk}\dfrac{\partial Q_{ij}}{\partial x_l}\dfrac{\partial Q_{ij}}{\partial x_k},\notag
\end{align}
where Einstein summation over repeated indices is implied and $\epsilon$ represents the Levi-Civita tensor. 

While the tensorial Landau-de Gennes framework na\-tu\-ra\-lly captures the complex phase behavior and defect topologies of nematic liquid crystals, its elastic constants ($L_i$) can be directly mapped to the classical Oseen-Frank (OF) deformations in the uniaxial limit.

To be more precise, $L_1$, $L_2$, and $L_3$ are linear combinations of the splay ($K_1$), twist ($K_2$), bend ($K_3$), and saddle-splay ($K_{24}$) moduli, and the connection between the LdG elastic constants and the OF ones being given by \Cref{eq:el_cst_LdG=OF} from the appendix.

Since we are interested in analysing a LdG elastic energy up to three elastic constants, we impose:
\begin{align}\label{eq:K1=K3_q0=0}
L_4=L_6=0\overset{\eqref{eq:el_cst_LdG=OF}}{\Longleftrightarrow} q_0=0\;\text{and}\;K_1=K_3.
\end{align}

Moreover, we recall that the elastic constants $L_1$, $L_2$ and $L_3$ verify the Ericksen inequalities, that is:
\begin{align}\label{eq:Ericksen}
L_1>0,\;-L_1<L_3<2L_1,\;-\dfrac{3}{5}L_1-\dfrac{1}{10}L_3<L_2.
\end{align}

% From a phenomenological standpoint, employing the one-constant approximation ($L_2=0, L_3=0$) imposes a strict physical constraint: it assumes that all fundamental deformations---splay, twist, and bend---incur at an identical energetic cost. Consequently, determining the precise threshold at which $L_2$ and $L_3$ can be neglected is equivalent to determining when the energetic discrepancies between splay, twist, and bend deformations become negligible compared to the total free energy of the confined system.

The bulk energy is given by
\begin{align*}
f_b(Q)&=-\dfrac{a}{2}\text{tr}(Q^2)-\dfrac{b}{3}\text{tr}(Q^3)+\dfrac{c}{4}\big(\text{tr}(Q^2)\big)^2,
\end{align*}
where $a,b,c\in(0,\infty)$ are material parameters, and we choose the Rapini-Papoular type of surface energy, which is given by
\begin{align*}
f_s(Q,\nu)=\dfrac{W}{2}\big|Q-Q_{\nu}\big|^2,
\end{align*}
where $W\in(0,\infty)$ represents the anchoring strength and $Q_{\nu}=s_0\bigg(\nu\otimes\nu-\dfrac{1}{3}I_3\bigg)$, with $\nu$ being the interior normal to the surface of the colloid $B(C_0,R)$.

\section{Main results}\label{sec:main_results}

Generalized Optimal Scaling (GOS), developed in \cite{rusconi25}, is an advanced, equation-free methodology designed to systematically reduce the complexity of physical models and effectively control over-parameterization. A key capability of this approach is the minimization of governing parameters, ensuring that a given model relies strictly on the minimal number of independent dimensionless parameters.

From a practical perspective, GOS builds upon the foundational features of prior scaling techniques by introducing a rigorous quantitative criterion for evaluating physical parameters. This explicit criterion enables us to systematically discard specific terms—namely, those associated with negligibly small parameters—from complex governing equations without significantly compromising the accuracy of the predictive solutions.

We are therefore able to find scale separation regimes for the elastic constants from the Landau-de Gennes model: we first analyse the reduction $(L_1,L_2,L_3)\leadsto(L_1,L_2,0)$, in which we apply GOS with the target $L_3$, and then the reduction $(L_1,L_2,0)\leadsto(L_1,0,0)$, where we apply GOS with the target $L_2$.

For more details related to applying GOS to the previously introduced setting, we recommend the reader to see \Cref{sec:GOS}.

At the same time, it is worth mentioning that if one wants to find scale separation regimes for the anchoring strength $W$, or the size of the lattice $L$, or the radius of the colloid $R$, or any other physical parameter used, then the regimes are presented in \Cref{tab:regimes_for_physical_param_3_cst}, for the $L_3\neq 0$ case, or \Cref{tab:regimes_for_physical_param_2_cst}, for the $L_3=0$ case.

We continue by presenting the main results from each case mentioned above.

\subsection{The $(L_1,L_2,L_3)\leadsto(L_1,L_2,0)$ case}

In this case, the threshold found for $L_3$ is given by the following inequality:
\begin{align}\label{eq:threshold_3cst}
    L_3<<\sqrt[41]{L_1^{15}\cdot L_2^{15}\cdot a\cdot b\cdot c\cdot W^8\cdot L^7\cdot R^7}:=\overline{L_3}.
\end{align}

We consider the specific values for all physical pa\-ra\-me\-ters in \Cref{sec:initial_param}. We then choose to study the following cases for the choice of $L_3$: 
\begin{itemize}
    \item $L_3=0$, which represents the case where we can neglect $L_3$ from our model,
    \item $L_3=\lambda_3\cdot \overline{L_3},\;\lambda_3=10^{-t},\;\forall t\in\{0,1,2,\ldots,7\}$.
\end{itemize}

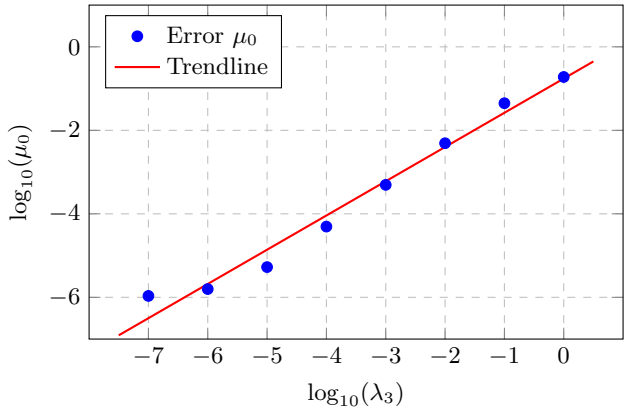
\begin{figure}[htbp]
\centering
\begin{tikzpicture}
\begin{axis}[
xlabel={$\log_{10}(\lambda_3)$},
ylabel={$\log_{10}(\mu_0)$},
xmin=-8, xmax=1,
ymin=-7, ymax=1,
xtick={-7,-6,-5,-4,-3,-2,-1,0},
grid=major,
grid style={dashed, gray!50},
legend pos=north west,
width=\columnwidth,
height=6cm,
legend cell align={left}
]

% Puncte Centered case
\addplot[
only marks,
mark=*,
color=blue
] coordinates {(-7,-5.9667) (-6,-5.8022) (-5,-5.2760) (-4,-4.3061) (-3,-3.3065) (-2,-2.3091) (-1,-1.3510) (0,-0.7219)};
\addlegendentry{Error $\mu_0$}

% Trendline Centered
\addplot[
domain=-7.5:0.5,
color=red,
thick
] {0.820*x - 0.760};
\addlegendentry{Trendline}

\end{axis}
\end{tikzpicture}
\caption[3cst]{%
\begin{tabular}[t]{@{}c@{}}
Errors for the $(L_1,L_2,L_3)\leadsto(L_1,L_2,0)$ case
%Trendline: $\log_{10}(\mu_0) = 0.820\log_{10}(\lambda_3) - 0.760$% \\
% Pearson coefficient: 0.9893
\end{tabular}%
}
\label{fig:3cst_error}
\end{figure}

Using \texttt{openQmin}, we find numerical solutions to each problem considered and we denote by:
\begin{itemize}
    \item $Q_{3,t}$ the solution for each $L_3=10^{-t}\cdot \overline{L_3}$, $t\in\overline{0,7}$,
    \item $Q_3$ the solution for $L_3=0$.
\end{itemize}
More details about utilising \texttt{openQmin} can be found in \Cref{sec:openQmin}. We compute then the discrete $L^2$ norm (denoted $L^2_d$) of each solution and measure the deviation from the simplified model $(L_3=0)$ by computing the following error:
\begin{align*}
    \mu_0=\dfrac{\|Q_{3,t}-Q_3\|_{L^2_d(\Omega)}}{\|Q_3\|_{L^2_d(\Omega)}}.
\end{align*}
For the numerical values of $\mu_0$ and $\lambda_3$, we pass them into the $\log-\log$ scale and the Pearson coefficient for the achieved data is $0.9893$, which tells us that we expect a linear dependency in this scale. The plot of all data points can be found in \Cref{fig:3cst_error} and we find the following relation:
\begin{align*}
    \log_{10}(\mu_0)\approx 0.820\log_{10}(\lambda_3)-0.760,
\end{align*}
which is equivalent to
\begin{align}\label{eq:error_3cst_approx}
    \mu_0\approx 0.174\cdot\lambda_3^{0.820}.
\end{align}

\subsection{The $(L_1,L_2,0)\leadsto(L_1,0,0)$ case} 

We proceed in a similar fashion for the case $(L_1,L_2,0)\leadsto(L_1,0,0)$ and we obtain the following threshold for $L_2$:
\begin{align}\label{eq:threshold_2cst}
    L_2<<\sqrt[26]{L_1^{15}\cdot a\cdot b\cdot c\cdot L^7\cdot R^7\cdot W^8}:=\overline{L_2}.
\end{align}
We choose once again
\begin{align*}
    L_2=\lambda_2\cdot \overline{L_2},\;\lambda_2=10^{-t},\;\forall t\in\{0,1,2,\ldots,7\}.
\end{align*}
and also the case $L_2=0$. 

Once again, the numerical data in the $\log-\log$ scale give us a Pearson coefficient of $0.9976$, so, once again, the plot can be found in \Cref{fig:2cst_error} and we find the following linear dependency in this scale:
\begin{align*}
    \log_{10}(\mu_0)\approx 0.928\log_{10}(\lambda_2)-0.034,
\end{align*}
which is equivalent to
\begin{align}\label{eq:error_2cst_approx}
    \mu_0\approx 0.925\cdot \lambda_2^{0.928}.
\end{align}

\begin{figure}[htbp]
\centering
\begin{tikzpicture}
\begin{axis}[
xlabel={$\log_{10}(\lambda_2)$},
ylabel={$\log_{10}(\mu_0)$},
xmin=-8, xmax=1,
ymin=-7, ymax=1,
xtick={-7,-6,-5,-4,-3,-2,-1,0},
grid=major,
grid style={dashed, gray!50},
legend pos=north west,
width=\columnwidth,
height=6cm,
legend cell align={left}
]

% Puncte Centered case
\addplot[
only marks,
mark=*,
color=blue
] coordinates {
(-7,-6.2316) (-6,-5.6961) (-5,-4.8607) (-4,-3.8672) (-3,-2.8675) (-2,-1.8790) (-1,-0.9931) (0,0.1288)
};
\addlegendentry{Error $\mu_0$}

% Trendline
\addplot[
domain=-7.5:0.5,
color=red,
thick
] {0.928*x - 0.034};
\addlegendentry{Trendline}

\end{axis}
\end{tikzpicture}
\caption[2cst]{%
\begin{tabular}[t]{@{}c@{}}
Errors for the $(L_1,L_2,0)\leadsto(L_1,0,0)$ case
%Trendline: $\log_{10}(\mu_0) = 0.928\log_{10}(\lambda_3) - 0.034$% \\
%Pearson coefficient: 0.9976
\end{tabular}%
}
\label{fig:2cst_error}
\end{figure}

We also highlight \Cref{tab:decision_process}, which illustrates the steps one should consider when analysing scale separation regimes using GOS.

\begin{table*}[htbp]
    \centering
    \caption{Decision process for using simplified models by GOS}
    \label{tab:decision_process}
    \begin{ruledtabular}
    \begin{tabular}{cl}
         Step 1: & Set the explicit values for all physical parameters involved ($L_1$, $L_2$, $L_3$, $a$, $b$, $c$, $W$, $R$ and $L$). \\
         \hline
         Step 2: & Choose the physical parameter $P$ for which we want to study scale separation regimes.\\
         \hline
         Step 3: & Determine and compute the critical threshold $\overline{P}$ using GOS \cite{rusconi25}.\\
         \hline
         Step 4: & Compare $P$ with $\overline{P}$ by evaluating $P/\overline{P}$.\\
         \hline
         Step 5: & If $P/\overline{P}\ll 1$, then the simplified model can be used.
    \end{tabular}
    \end{ruledtabular}
\end{table*}

\begin{remark}
In our case, we were able to find an additional information: for the three elastic constant case with the target $p=3$ or for the two elastic constant case with the target $p=2$, if $\lambda_p<<1$, then the simplified model can be used with an error of the same order of magnitude with $\lambda_p$.
\end{remark}

\begin{remark}
Our study illustrates one possible way in which one could reduce the 3-elastic constant model from the Landau-de Gennes to the widely used one constant approximation. If one starts with $(L_1,L_2,L_3)$, then \Cref{eq:threshold_3cst} establishes whether $L_3$ can be dropped or not. If so, then the $(L_1,L_2,0)$ is being considered, where, once again, by using GOS and \Cref{eq:threshold_2cst}, we can establish whether $L_2$ can be dropped or not. If both cases are true, one ends with a justification for the use of the one-constant approximation.
\end{remark}

\section{Initial physical parameters}\label{sec:initial_param}

We consider $L=3\times 10^{-7}\;\text{m}=300\;\text{nm}$. We are aware that this scale does not highlight the common behaviour of nematic liquid crystals used in devices, where larger samples of dimensions $\approx 10\;\mu$m are used. However, we are interested in analysing small variations of elastic con\-stants, hence a more physical relevant sample, with size $300$ nm, is suited. Let the radius of the colloid to be $R=L/4=75$ nm, such that its diameter is half of the size of the container.

For the bulk coefficients, we consider, as in \cite{ravnik2009}:
\begin{align}\label{eq:A_B_C_Ravnik}
    a&=0.172\times 10^6\;\text{J}/\text{m}^3,\notag\\
    b&=2.120\times 10^6\;\text{J}/\text{m}^3,\\
    c&=1.730\times 10^6\;\text{J}/\text{m}^3,\notag
\end{align}
where we recall that $1\;\text{J}=1\;\text{kg}\cdot\text{m}^2\cdot\text{s}^{-2}$.

\begin{remark}
    The choices for the bulk energy coefficients from \Cref{eq:A_B_C_Ravnik} are taken from \cite{ravnik2009} and they represent the values for the 5CB material.
\end{remark}

The bulk coefficients from \Cref{eq:A_B_C_Ravnik} give us the following (dimensionless) nematic preferred degree of order:
\begin{align*}
    S=\dfrac{-b+\sqrt{b^2+24ac}}{6c}\approx 0.533
\end{align*}
and we denote
\begin{align}\label{eq:alfa}
    \alpha=\dfrac{4}{9S^2}\approx 1.565,
\end{align}
and, together with \Cref{eq:K1=K3_q0=0,eq:el_cst_LdG=OF}, we obtain:
\begin{align}\label{eq:simplified_L1_L2_L3_L4_L6}
\begin{cases}
    L_1&=\dfrac{\alpha}{2}K_2,\\
    L_2&=\alpha(K_1-K_{24}),\\
    L_3&=\alpha(K_{24}-K_2),\\
    L_4&=0,\\
    L_6&=0.
\end{cases}
\end{align}

By inverting the previous equations, we obtain:
\begin{align}\label{eq:el_cst_OF=LdG}
\begin{cases}
K_1=K_3&=\dfrac{2L_1+L_2+L_3}{\alpha}\\ \vspace{2mm}
K_2&=\dfrac{2L_1}{\alpha}\\ \vspace{2mm}
K_{24}&=\dfrac{2L_1+L_3}{\alpha}.
\end{cases}
\end{align}

We desire to use in our numerical simulations strictly positive elastic constants: $L_1>0$, $L_2\geq 0$ and $L_3\geq 0$. Using this constraint and \Cref{eq:simplified_L1_L2_L3_L4_L6}, the Ericksen inequalities from \Cref{eq:Ericksen} become:
\begin{align*}
    L_1>0\;&\;\Rightarrow K_2>0,\\
    L_2\geq 0\;&\;\Rightarrow K_1\geq K_{24},\\
    L_3\geq 0\;&\;\Rightarrow K_{24}\geq K_2,\\
    -L_1<L_3<2L_1\;&\;\Rightarrow K_{24}<2K_2,
\end{align*}
where we have used also that $\alpha>0$ (see \Cref{eq:alfa}), and the last inequality from \Cref{eq:Ericksen} is automatically satisfied, since we considered $L_1>0$, $L_2\geq 0$ and $L_3\geq 0$. We are therefore interested in considering OF elastic constants such that:
\begin{align*}
    0<K_2\leq K_{24}\leq K_1=K_3\;\text{and}\;K_{24}<2K_2.
\end{align*}

\begin{remark}
    In \cite{bankova24}, there are presented the following measured numerical values for $K_1$, $K_2$ and $K_3$ for the $5CB$ material: $K_1\approx 5.6$ pN, $K_2\approx 3.7$ pN and $K_3\approx 8.2$ pN. We first decided to use integer values for the OF elastic constants. Then, since we are interested in choosing $K_1=K_3$ large enough such that we still have plenty of options for integer va\-lues for $K_2$ and $K_{24}$, we decided to use $K_1=K_3=8$ pN. For $K_2$, one could of course consider the rounding of the approximation from \cite{bankova24} or even the exact approximation, but here we decided to choose $K_2=5$ pN, simply to avoid any obvious ratios between $K_1$ and $K_2$ which might lead to very particular behaviours.
\end{remark}

In all cases studied, we consider 
\begin{align*}
    K_1=K_3=8\;\text{pN and}\;K_2=5\;\text{pN}.
\end{align*}
For $K_{24}$, we consider two cases:
\begin{align*}
    K_{24}&=7\;\text{pN for}\;(L_1,L_2,L_3)\leadsto(L_1,L_2,0)
\end{align*}
and
\begin{align*}
    K_{24}&=5\;\text{pN for}\;(L_1,L_2,0)\leadsto(L_1,0,0).
\end{align*}

The previous choices give us the following elastic constants from the LdG model:
\begin{align}\label{eq:init_elastic_3cst}
    L_1\approx 3.913\;\text{pN},\; L_2\approx 1.565\;\text{pN},\;L_3\approx 3.131\;\text{pN},
\end{align}
for the $(L_1,L_2,L_3)\leadsto(L_1,L_2,0)$ case, and
\begin{align}\label{eq:init_elastic_2cst}
    L_1\approx 3.913\;\text{pN},\; L_2\approx 1.565\;\text{pN},\;L_3=0\;\text{pN},
\end{align}
for the $(L_1,L_2,0)\leadsto(L_1,0,0)$ case.

We also consider the case of strong anchoring strength with 
\begin{align*}
W=5\cdot 10^{-4}\;\text{J}/\text{m}^2. 
\end{align*}

Having set all the initial parameters, then the critical elastic constants presented in \Cref{sec:main_results} become:
\begin{align*}
    \overline{L_3}\approx 9.276\;\text{pN}
\end{align*}
and
\begin{align*}
    \overline{L_2}\approx 25.897\;\text{pN}.
\end{align*}

\begin{remark}
    In the first physical setting, we see that:
    \begin{align*}
        \lambda_3=\dfrac{L_3}{\overline{L_3}}\approx 0.338.
    \end{align*}
    If $Q_{3,init}$ is the numerical minimizer of \Cref{eq:min_problem} in which we use $L_3$, then, using \Cref{eq:error_3cst_approx}, we obtain that:
    \begin{align*}
        \dfrac{\|Q_{3,init}-Q_{3}\|_{L_d^2(\Omega)}}{\|Q_{3}\|_{L^2_d(\Omega)}}\approx 0.174\cdot 0.338^{0.820}\approx 0.071,
    \end{align*}
    where $Q_3$ is the solution for the $(L_1,L_2,0)$ case. The previous approximation tells us that we can choose the simplified model with an error of approximate $7\%$ of the norm of the solution of the simplified model.
\end{remark}

\begin{remark}
    In the second physical setting, we have:
    \begin{align*}
        \lambda_2=\dfrac{L_2}{\overline{L_2}}\approx 0.060.
    \end{align*}
    Using the same notation as before for the two elastic constants case, we get, using \Cref{eq:error_2cst_approx}, the following:
    \begin{align*}
        \dfrac{\|Q_{2,init}-Q_{2}\|_{L^2_d(\Omega)}}{\|Q_{2}\|_{L^2_d(\Omega)}}\approx 0.925\cdot 0.060^{0.928}\approx 0.068,
    \end{align*}
    where $Q_2$ is the solution for the $(L_1,0,0)$ case. We see that the previous approximation gives us the same deviation of $7\%$ as in the previous case.
\end{remark}

\begin{remark}
    The initial physical parameters that we have allow us to consider the following situation: we start with $L_1$, $L_2$ and $L_3$ given in \Cref{eq:init_elastic_3cst}. We have applied GOS and we have seen that our target solution, $Q_{3,init}$, has the following property:
    \begin{align*}
        0.93 \leq\dfrac{\|Q_{3,init}\|_{L^2_d(\Omega)}}{\|Q_{3}\|_{L^2_d(\Omega)}}\leq 1.07
    \end{align*}
    But $Q_{3}$ is exactly $Q_{2,init}$, since they are the numerical solutions of the case $(L_1,L_2,0)$, with $L_1$ and $L_2$ given in \Cref{eq:init_elastic_2cst}, so one could apply GOS once again and get that:
    \begin{align*}
        0.86\leq \dfrac{\|Q_{3,init}\|_{L^2_d(\Omega)}}{\|Q_{2}\|_{L^2_d(\Omega)}}\leq 1.14.
    \end{align*}
    At the same time, $Q_{2}$ represents the numerical solution for the one-constant approximation $(L_1,0,0)$. 

    Hence, in this given context, one could apply GOS twice:
    \begin{align*}
        (L_1,L_2,L_3)\leadsto(L_1,L_2,0)\leadsto (L_1,0,0),
    \end{align*}
    to justify using the widely-used one-constant approximation with an error less than $14\%$, instead of the initial 3-constant model.
\end{remark}

% Having all of these initial parameters, the critical values presented in the previous section become:
% \begin{align}
%     C_3\approx 5.993
% \end{align}
% and
% \begin{align}
%     C_2\approx 16.730.
% \end{align}
% One important aspect here is that the constants $C_2$ and $C_3$ are independent of rescalings (see [Proposition Appendix]).

\section{Computational methodology}\label{sec:openQmin}

To quantitatively evaluate the limits of the one-constant and two-constant elastic energy approximations, we performed three-dimensional numerical simulations using \texttt{open-Qmin}, an open-source computational framework specifically designed for nematic liquid crystals. Developed by Sussman and Beller (see \cite{sussman19} for more details), \texttt{open-Qmin} relies on a finite-difference lattice discretization of the continuum Landau-de Gennes free energy functional. The software evaluates elastic, bulk and surface energy contributions, making it highly suitable for modeling confined geometries with topological defects.
\begin{figure}[htbp]
    \centering
    \includegraphics[width=\linewidth]{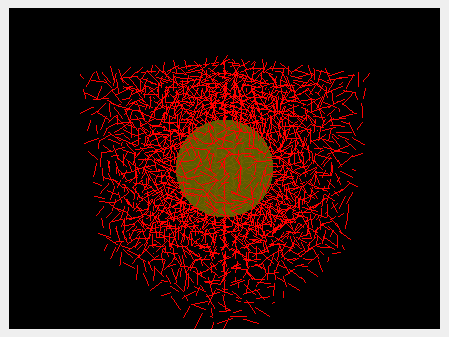}
    \caption{Initial random configuration in \texttt{open-Qmin}}
    \label{fig:open-Qmin-init-config}
\end{figure}

Rather than solving complex differential equations analytically, \texttt{open-Qmin} utilizes the Fast Inertial Relaxation Engine (FIRE) algorithm to minimize the total free energy of the system. Starting from an initial configuration of the tensorial order parameter $Q$ across the simulation lattice, as shown in \Cref{fig:open-Qmin-init-config}, the FIRE method acts as an efficient gradient-descent minimizer, iteratively relaxing the $Q$-tensor field until the net thermodynamic force at each lattice site approaches zero. This process yields the equilibrium director configuration and precisely resolves the spatial distribution of the nematic degree of order, naturally capturing defect structures where the orientational order breaks down. Crucially for our study, \texttt{open-Qmin} permits the explicit and independent definition of the multi-constant Landau-de Gennes elastic coefficients ($L_1$, $L_2$, $L_3$), enabling a direct numerical comparison between the full multi-constant models and their simplified one- or two-constant counterparts.

Each run of \texttt{open-Qmin} generates a text file, which includes the following information:
\begin{itemize}
    \item the position of a point from the lattice - $(x_0,y_0,z_0)$;
    \item the five components of $Q$ from that point - $q_1$, $q_2$, $q_3$, $q_4$ and $q_5$;
    \item the type of material at that point: $0$ for nematic liquid crystal particles and $1$ for colloids;
    \item the nematic preferred degree of order at that point, which is the largest eigenvalue of $Q(x_0,y_0,z_0)$ (not relevant for our study).
\end{itemize}

This allows to compute a discrete $L^2$ norm of the simulated $Q$-tensor $Q_s$, denoted $\|Q_s\|_{L^2_d(\Omega)}$ for the chosen configuration:
\begin{align*}
    \|Q_s\|_{L^2_d(\Omega)}=\sum_{i,j,k=1}^{N}(1-t(i,j,k))\cdot\text{tr}(Q_s^2(i,j,k)),
\end{align*}
where $N$ represents the number of points considered for each side of the cubic lattice and $t(i,j,k)$ represents whether the material at the point $(i,j,k)$ is nematic liquid crystal (then it is $0$) or just part of the colloid (when it is $1$).

We assume that, for $N$ large enough, we have:
\begin{align*}
    \|Q_s\|_{L^2_d(\Omega)}\approx \|Q\|_{L^2(\Omega)},
\end{align*}
where $Q$ represents the minimizer of the free energy functional considered in \Cref{eq:min_problem}.

One important aspect worth mentioning is the fact that \texttt{open-Qmin} is automatically rescaling all the inputs such that the leading coefficient from the bulk energy, $a$, is of module $1$. In order to avoid such automatic rescaling, we manually rescale all of our parameters before inserting them in \texttt{open-Qmin}.

\begin{figure}[htpb]
    \centering
    \includegraphics[width=\linewidth]{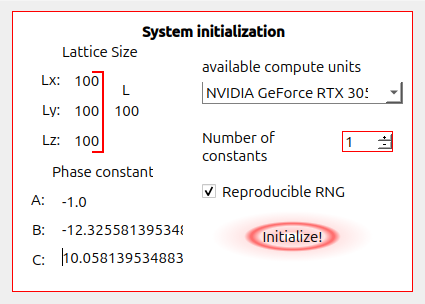}
    \caption{\texttt{open-Qmin} - Setting $L$, $a$, $b$ and $c$}
    \label{fig:open-Qmin-open-1st-page}
\end{figure}

At the same time, there is another reason for manually rescaling parameters: the size of the lattice. According to \cite{sussman19}, when operating with single-core \texttt{GPUs}, it is recommended to have a lattice size smaller than $300$, as the input in \texttt{open-Qmin} (shown in \Cref{fig:open-Qmin-open-1st-page}). 

In our case, the \texttt{GPU} is represented by a single unit of \texttt{NVIDIA GeForce RTX 3050} (the mobile version for laptops), with an internal memory of \texttt{6 GB}. Due to the internal memory of the \texttt{GPU}, we can not achieve lattice sizes larger than $150$, hence, we have run all simulations with a lattice size of $100$. 

We therefore need to apply two rescalings: one change of variables such that the $300$ nm length is scaled to $100$ and a second one for the energy such that $|a|$ becomes $1$.

\begin{figure}[htbp]
    \centering
    \includegraphics[width=\linewidth]{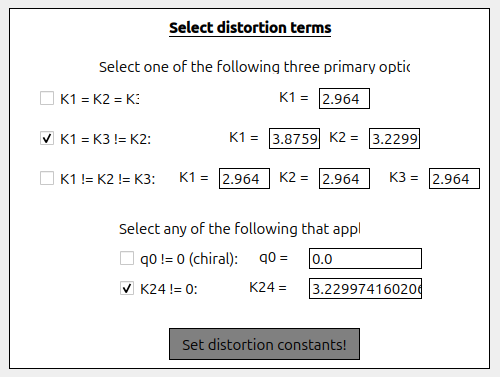}
    \caption{\texttt{open-Qmin} - Setting the elastic constants}
    \label{fig:open-Qmin-open-2nd-page}
\end{figure}

Moreover, since \texttt{open-Qmin} invites the user to insert the Oseen-Frank elastic constants $K_1$, $K_2$, $K_3$ and $K_{24}$ (see \Cref{fig:open-Qmin-open-2nd-page}), we only present in this section  our choices for these parameters, instead of $L_1$, $L_2$ and $L_3$, since they can be reconstructed using \Cref{eq:el_cst_LdG=OF}.

We transform all initial physical parameters such that their units of measure are written with respect to kilograms, meters and seconds. Then our initial parameters assume the following numerical values in such a system of units:
\begin{align*}
    \begin{cases}
        K_1=K_3=8\cdot 10^{-12},\\
        K_2=5\cdot 10^{-12},\\
        a=0.172\cdot 10^6,\\
        b=2.120\cdot 10^6,\\
        c=1.730\cdot 10^6,\\
        W=5\cdot 10^{-4},\\
        L=3\cdot 10^{-7},\\
        R=0.75\cdot 10^{-7}
    \end{cases}
\end{align*}
and
\begin{align*}
    K_{24}=\begin{cases}
        7\cdot 10^{-12},\;\text{for}\;(L_1,L_2,L_3)\leadsto(L_1,L_2,0),\\
        5\cdot 10^{-12},\;\text{for}\;(L_1,L_2,0)\leadsto(L_1,0,0).
    \end{cases}
\end{align*}

After the change of variables $\widetilde{x}=\dfrac{x}{\theta_1}$ in \Cref{eq:elastic_energy_LdG}, where $\theta_1=3\cdot 10^{-9}$, we get the following rescaled pa\-ra\-me\-ters, denoted now with tilde:
\begin{align*}
    \begin{cases}
        \widetilde{K_1}=\widetilde{K_3}=24\cdot 10^{-21},\\
        \widetilde{K_2}=15\cdot 10^{-21},\\
        \widetilde{a}=4.644\cdot 10^{-21},\\
        \widetilde{b}=57.24\cdot 10^{-21},\\
        \widetilde{c}=46.71\cdot 10^{-21},\\
        \widetilde{W}=45\cdot 10^{-22},\\
        \widetilde{L}=100,\\
        \widetilde{R}=25
    \end{cases}
\end{align*}
and
\begin{align*}
    \widetilde{K_{24}}=\begin{cases}
        21\cdot 10^{-21},\;\text{for}\;(L_1,L_2,L_3)\leadsto(L_1,L_2,0),\\
        15\cdot 10^{-21},\;\text{for}\;(L_1,L_2,0)\leadsto(L_1,0,0),
    \end{cases}
\end{align*}
where we have used \Cref{eq:change_of_variables}. In this way, we have transformed our parameters for the \texttt{open-Qmin} setting: the lattice size of $100$. However, since $|\widetilde{a}|\neq 1$, we now rescale \Cref{eq:elastic_energy_LdG} with $\theta_2=\widetilde{a}$. Using once again \Cref{eq:change_of_variables}, we get the actual input data for the \texttt{open-Qmin} simulations:
\begin{align}\label{eq:actual_input_open-Qmin_1}
    \begin{cases}
        \texttt{K}_1=\texttt{K}_3\approx 5.16795865633075,\\
        \texttt{K}_2\approx 3.22997416020672,\\
        \texttt{a}=1,\\
        \texttt{b}\approx 12.3255813953488,\\
        \texttt{c}\approx 10.0581395348837,\\
        \texttt{W}\approx 0.968992248062015,\\
        \texttt{L}=100,\\
        \texttt{R}=25
    \end{cases}
\end{align}
and
\begin{align}\label{eq:actual_input_open-Qmin_2}
\texttt{K}_{24}\approx\begin{cases}
4.5219638242894,\;\text{for}\;(L_1,L_2,L_3)\leadsto(L_1,L_2,0)\\
3.22997416020672,\;\text{for}\;(L_1,L_2,0)\leadsto(L_1,0,0).
\end{cases}
\end{align}
In this way, \Cref{eq:actual_input_open-Qmin_1} and \Cref{eq:actual_input_open-Qmin_2} represent the initial physical parameters, presented in \Cref{sec:initial_param}, transformed and prepared for simulations using \texttt{open-Qmin}, for our specific \texttt{GPU} constraints. We decided to leave here all decimals used for our simulations, for the ease of replication.

If one wants to see or recreate our numerical simulations, depending on our choices of $\lambda_2$ and $\lambda_3$, then one could check \Cref{sec:numerical_data} (more specifically, \Cref{tab:numerical_data_3cst,tab:numerical_data_2cst}), but also our Zenodo repository \cite{zenodo}.

\begin{remark}
    One option that \texttt{open-Qmin} can use the \textit{Reproducible RNG} function. This keeps the initial random configuration, presented in \Cref{fig:open-Qmin-init-config}, whenever the system is being reset in order to generate a new numerical simulation. This initial random configuration might depend locally on the machine that is being used, hence, there might be a chance that the numerical data achieved for the errors from \Cref{sec:main_results} to be slightly different.
\end{remark}

% \begin{remark}
%     Another option that \texttt{open-Qmin} can use is the \textit{force cut-off error} for the FIRE algorithm. By default, this is set to $10^{-12}$ and it is an instrument for adjusting a precision level for the numerical solution of the configuration. In order to measure such small deviations for the $Q$-tensor solutions, we have left the standard threshold of $10^{-12}$, since we make modifications up to the $7^{\text{th}}$ decimal of the elastic constants $L_2$ and $L_3$. However, modifying this can heavily impact the amount of time consumed for each simulation and, at the same time, if one uses lower precisions for the solutions (larger force cut-off errors), then, of course, also the errors $\mu_0$ from \Cref{sec:main_results} might modify heavily. 
% \end{remark}

\begin{remark}
An essential computational parameter within the \texttt{open-Qmin} framework is the force cut-off tolerance for the FIRE algorithm, which dictates the convergence criteria for the energy minimization. By default, this is set to $10^{-12}$. In order to accurately capture the subtle structural deviations in the $Q$-tensor---particularly since we investigate perturbations up to the $7^{\text{th}}$ decimal of the elastic constants $L_2$ and $L_3$---we strictly maintained this high-precision threshold. It is crucial to note that the measured deviation $\mu_0$ is highly sensitive to this cut-off parameter. If a lower precision (e.g., a larger cut-off error such as $10^{-10}$) is employed, the numerical noise of early convergence artificially inflates $\mu_0$ by one to two orders of magnitude. Under such conditions, the numerical error overshadows the physical model error, inherently altering the scaling coefficients in the log-log regime (we still keep the same linear dependency, but with different coefficients). Therefore, maintaining a strict force cut-off tolerance of $10^{-12}$ might be seen as a computational preference, but at the same time, it cleanly decouples the physical scaling of the elastic energy from the underlying algorithmic noise.
\end{remark}

Once the FIRE algorithm reaches the prescribed force cut-off tolerance, the final equilibrium configuration can be exported as a text file. As previously detailed, our simulations utilize a lattice size of 100, resulting in one million data points ($100^3$) per file. Because each lattice point stores seven distinct numerical values---five for the $Q$-tensor components, one material identifier indicating whether the point is a colloid or nematic liquid crystal, and one for the maximum eigenvalue---a single simulation yields seven million data entries and produces a file of approximately 68 MB. Standard spreadsheet applications are typically ill-equipped to handle datasets of this magnitude. Therefore, we processed the raw data using RowZero, a high-performance online data platform. This environment allowed us to efficiently calculate the discrete $L^2$ norms of the configurations and compute the component-wise differences between distinct simulations to determine our error metrics.

Fully processed data sheets exported from RowZero reach sizes of around 450 MB per simulation. To keep our shared data repository accessible and practical for download, we have chosen not to distribute these massive processed sheets. Instead, we provide the raw 68 MB text files directly generated by \texttt{open-Qmin}. By sharing these foundational files, independent researchers can readily reconstruct the numerical data and recompute the discrete norms to verify the deviations using their own preferred computational tools.

\begin{figure}[htbp]
    \centering
    \includegraphics[width=0.4\textwidth]{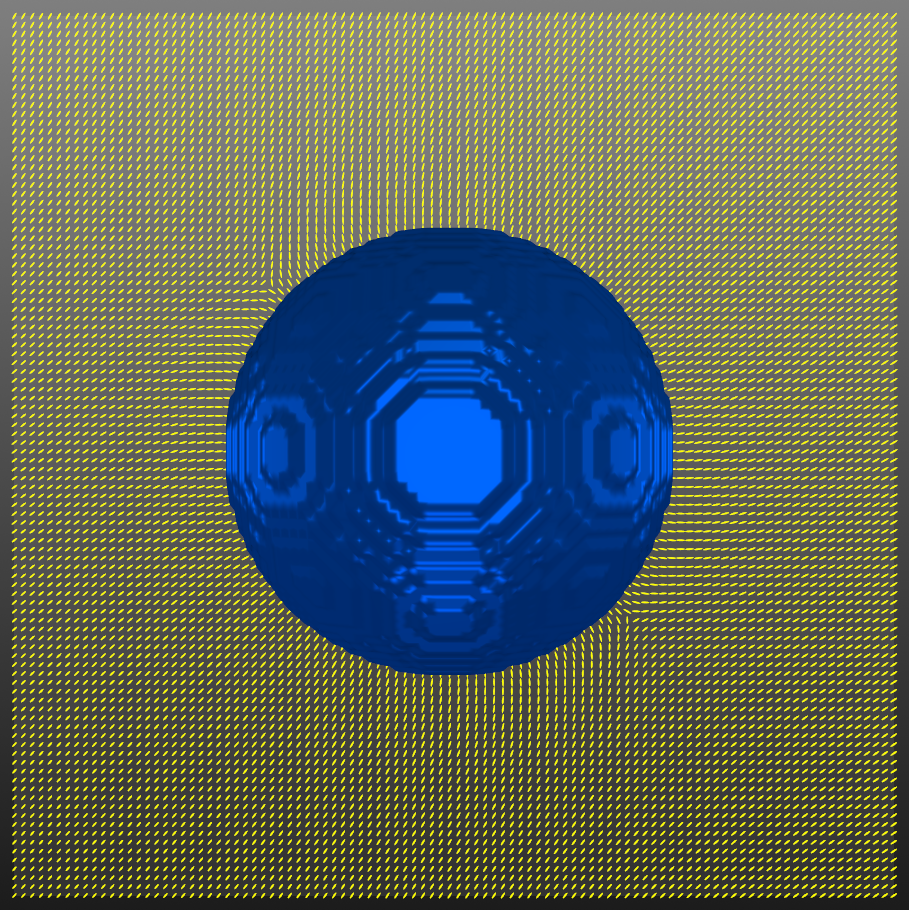}
    \caption{Visual representation of $Q_{3}$ using \texttt{open-Viewmin}}
    \label{fig:viewmin_Q3}
\end{figure}

Beyond quantitative data analysis, the qualitative features of the nematic configurations can be visualized using \texttt{open-Viewmin}. Also developed by Sussman and Beller, this companion application utilizes Python scripts to render detailed, three-dimensional visual representations of the simulated $Q$-tensor fields. In \Cref{fig:viewmin_Q3}, we see the representation of the solution $Q_3$ (computed for $\lambda_3=0$ -- which implies $L_3=0$), in the $yz$ plane.

\begin{figure*}[htbp]
\centering
\begin{subfigure}[htbp]{0.45\textwidth}
    \centering
    \includegraphics[width=\textwidth]{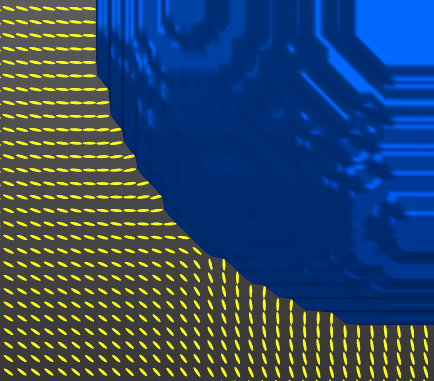}
    \caption{$\lambda_3=1$}
    \label{fig:viewmin_lambda3_minus_0}
\end{subfigure}\;\;\;\;\;\;\begin{subfigure}[htbp]{0.45\textwidth}
    \centering
    \includegraphics[width=\textwidth]{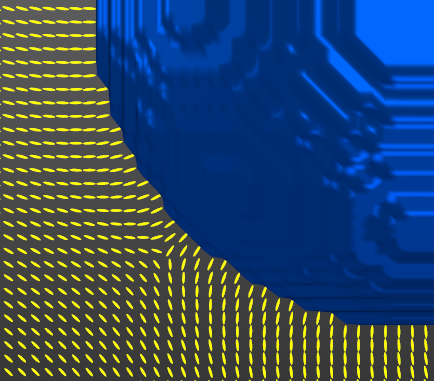}
    \caption{$\lambda_3=0.1$}
    \label{fig:viewmin_lambda3_minus_1}
\end{subfigure}

\begin{subfigure}[htbp]{0.45\textwidth}
    \centering
    \includegraphics[width=\textwidth]{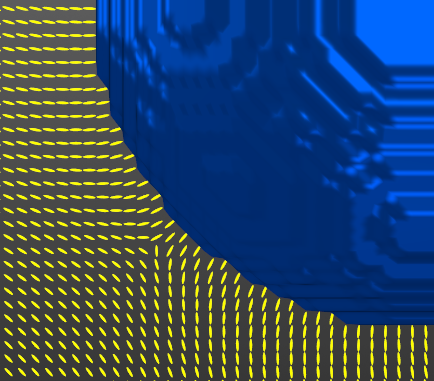}
    \caption{$\lambda_3=0.01$}
    \label{fig:viewmin_lambda3_minus_2}
\end{subfigure}\;\;\;\;\;\;\begin{subfigure}[htbp]{0.45\textwidth}
    \centering
    \includegraphics[width=\textwidth]{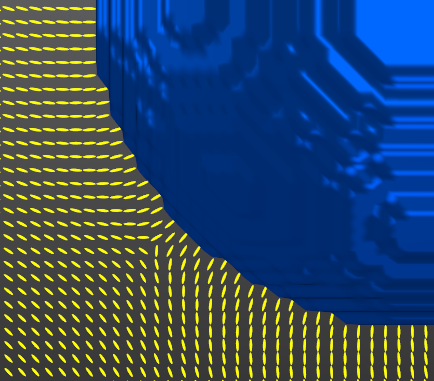}
    \caption{$\lambda_3=0$}
    \label{fig:viewmin_lambda3_0}
\end{subfigure}

\caption{Visual representations of four cases from the $(L_1,L_2,L_3)\leadsto (L_1,L_2,0)$ case}
\label{fig:viewmin_3cst_4cases_all}
\end{figure*}

The numerical data we have achieved for the errors $\mu_0$ in all studied cases from \Cref{sec:main_results} can be visually expressed using \texttt{open-Viewmin}. However, since we are studying variations of $10^{-7}$ in the elastic constants, the visual representation of such changes, even for the solutions, is non-significant. We decide to highlight in  only four such cases: $\lambda_3\in\{0,0.01,0.1,1\}$, since for all other intermediate values we are not able to see significant changes. Moreover, even for these four cases, we decided to zoom in exactly in the area where there are visible changes. If we analyse \Cref{fig:viewmin_lambda3_minus_0}, which is corresponding for the critical case $(L_1,L_2,L_3)$, with $L_3\neq 0$, we see that at some points on the colloid, the surface energy is not able to dominate the elastic one. However, as soon as $L_3$ becomes smaller and smaller, the surface energy starts forcing the particles to align perpendicularly to the colloidal particle, as seen in \Cref{fig:viewmin_lambda3_minus_1,fig:viewmin_lambda3_minus_2,fig:viewmin_lambda3_0}.

\section{Conclusions and Future Work}

In this study, we have provided a rigorous, quantitative framework for reducing the complexity of the multi-constant Landau-de Gennes elastic energy. By leveraging the Generalized Optimal Scaling (GOS) methodology, we derived explicitly computable parameter thresholds that determine when the three-constant elastic model $(L_1, L_2, L_3)$ can be safely simplified to a two-constant $(L_1, L_2, 0)$ model, and subsequently to the widely used one-constant $(L_1, 0, 0)$ approximation. Through comprehensive three-dimensional simulations using the \texttt{open-Qmin} framework, we numerically validated these theoretical bounds. Our computational results confirmed that as long as the neglected elastic constants fall below the critical thresholds established by GOS, the discrete $L^2$ deviation between the full and reduced nematic configurations remains strictly bounded and predictable. Ultimately, this provides a practical diagnostic tool for computational physicists to rigorously justify the use of simplified elastic models without introducing unphysical artifacts.

Building upon this foundational framework, there are several promising avenues for future research. From a physical modeling perspective, the GOS procedure can be extended to analyze nematic systems containing multiple colloidal particles of varying sizes. Additionally, we plan to investigate scale separation regimes for other cri\-ti\-cal parameters, such as the surface anchoring strength $W$, and examine how imposing distinct physical boundaries, such as rigid walls, alters the established scaling thresholds.

\begin{acknowledgments}
A.Z has been partially supported by the Basque Government through the BERC 2022-2025 program and by the Spanish State Research Agency through BCAM Severo Ochoa CEX2021-001142-S and through project PID2023-146764NBI00 funded by MICIU/AEI/10.13039/501100011033 and cofunded by the European Union. A.Z. was also partially supported by a grant of the Ministry of Research, Innovation and Digitization, CNCS - UEFISCDI, project number PN-III-P4-PCE-2021-0921, within PNCDI III.
\end{acknowledgments}

\section*{Data Availability Statement}

The raw Q-tensor components from the open-Qmin simulations and also the input parameters are available in the Zenodo repository \cite{zenodo}.

\begin{appendix}

\section{Connection between Oseen-Frank and Landau-de Gennes elastic energies}\label{sec:appendix_OF-LDG-connection}

The application openQmin allows the user to input values for the Oseen-Frank elastic constants. The Oseen-Frank elastic energy is:
\begin{align*}
f_{OF}&=\dfrac{1}{2}\bigg\{K_1\big(\nabla\cdot\hat{n}\big)^2+K_2\Big(\hat{n}\cdot\big(\nabla\times\hat{n}\big)+q_0\Big)^2+\\
&+K_3\Big|\big(\hat{n}\cdot\nabla)\hat{n}\Big|^2+K_{24}\nabla\cdot\Big[\big(\hat{n}\cdot\nabla\big)\hat{n}-\hat{n}\big(\nabla\cdot\hat{n}\big)\Big]\bigg\},
\end{align*}
where $K_1$, $K_2$, $K_3$ and $K_{24}$ are the Oseen-Frank elastic constants, representing each different types of elastic deformations: splay, twist, bend and saddle-splay. 

To recover the Landau-de Gennes elastic constants in \Cref{eq:elastic_energy_LdG} from the Oseen-Frank formulation, one could use the following formulas:
\begin{align}\label{eq:el_cst_LdG=OF}
L_1&=\dfrac{2}{27S^2}\big(K_3-K_1+3K_{2}\big),\notag\\
L_2&=\dfrac{4}{9S^2}\big(K_1-K_{24}\big),\notag\\
L_3&=\dfrac{4}{9S^2}\big(K_{24}-K_2\big),\\
L_4&=-\dfrac{8}{9S^2}q_0K_2\notag\\
&\text{and}\notag\\
L_6&=\dfrac{4}{27S^3}(K_3-K_1).\notag
\end{align}

\section{Scale separation using GOS}\label{sec:GOS}

\subsection{The $(L_1,L_2,L_3)\leadsto(L_1,L_2,0)$ case}

Following the same notation as in \cite{rusconi25}, we have 9 physical parameters in our model:
\begin{align*}
\overrightarrow{p}:=(L_1,L_2,L_3,a,b,c,W,L,R)^T\in \mathbb{R}^{N_p},\;N_p=9
\end{align*}
and they have the following units of measure:
\begin{align*}
\begin{cases}
\;[L_1]=[L_2]=[L_3]=\text{kg}\cdot\text{m}\cdot\text{s}^{-2},\\
\;[W]=\text{kg}\cdot\text{s}^{-2},\\
\;[a]=[b]=[c]=\text{kg}\cdot\text{m}^{-1}\cdot\text{s}^{-2},\\
\;[L]=[R]=\text{m}.
\end{cases}
\end{align*}

Then the matrix $\textbf{M}\in\mathcal{M}_{3\times 9}(\mathbb{R})$, containing all the exponents of the physical parameters with respect to the unit of measure kg, m and s, becomes:
\begin{align*}
\textbf{M}=\begin{pmatrix}
1&1&1&1&1&1&1&0&0\\
1&1&1&-1&-1&-1&0&1&1\\
-2&-2&-2&-2&-2&-2&-2&0&0
\end{pmatrix}.
\end{align*}
Then $\dim(\ker(\textbf{M}))=7:=N_0$ and a basis in $\ker(\textbf{M})$ is
\begin{align*}
\textbf{Z}=\begin{pmatrix}
-\dfrac{1}{2} & -\dfrac{1}{2} & -\dfrac{1}{2} & 0 & 0 & -1 & -1\\
0 & 0 & 0 & 0 & 0 & 0 & 1\\
0 & 0 & 0 & 0 & 0 & 1 & 0\\
\dfrac{1}{2} & \dfrac{1}{2} & -\dfrac{1}{2} & -1 & -1 & 0 & 0\\
0 & 0 & 0 & 0 & 1 & 0 & 0\\
0 & 0 & 0 & 1 & 0 & 0 & 0\\
0 & 0 & 1 & 0 & 0 & 0 & 0\\
0 & 1 & 0 & 0 & 0 & 0 & 0\\
1 & 0 & 0 & 0 & 0 & 0 & 0
\end{pmatrix},
\end{align*}
where the elements of the basis are written on the columns. The components of each vector give us the exponents that we use for the physical parameters to gen\-er\-ate $N_0=7$ corresponding governing parameters, according to \cite[Theorems 1 and 2]{rusconi25}, which are: 
\begin{align*}
    &\pi_1=\dfrac{R\sqrt{a}}{\sqrt{L_1}},\;\pi_2=\dfrac{L\sqrt{a}}{\sqrt{L_2}},\;\pi_3=\dfrac{W}{\sqrt{a\cdot L_3}},\\
    &\pi_4=\dfrac{c}{a},\;\pi_5=\dfrac{b}{a},\;\pi_6=\dfrac{L_3}{L_1},\;\pi_7=\dfrac{L_2}{L_1}.
\end{align*}
We can observe that all $\pi_i$, $i\in\{1,2,\ldots,7\}$ are dimensionless.

In our model, we have two characteristic constants, since we can only rescale the domain or the energy. We denote the characteristic constants by $\overrightarrow{\theta}=(\theta_1,\theta_2)^T$, where $N_X=2$, $[\theta_1]=\text{m}$ and $[\theta_2]=\text{kg}\cdot\text{m}^2\cdot\text{s}^{-2}$. Then we consider the change of variables
\begin{align*}
\overline{x}=\dfrac{x}{\theta_1}\;\text{and}\;\overline{\mathcal{F}}=\dfrac{\mathcal{F}}{\theta_2}.
\end{align*}

Using the form of our free energy functional from \Cref{eq:min_problem}, we find the following dimensionless coefficients for the rescaled problem are $\overline{\lambda}=(\lambda_1,\lambda_2,\ldots,\lambda_{N_d})\in\mathbb{R}^{N_d}$, with $N_d=N_p=9$:
\begin{align}\label{eq:change_of_variables}
\begin{cases}
\lambda_1=L_1\cdot\theta_1\cdot\theta_2^{-1},\\
\lambda_2=L_2\cdot\theta_1\cdot\theta_2^{-1},\\
\lambda_3=L_3\cdot\theta_1\cdot\theta_2^{-1},\\
\lambda_4=a\cdot\theta_1^3\cdot\theta_2^{-1},\\
\lambda_5=b\cdot\theta_1^3\cdot\theta_2^{-1},\\
\lambda_6=c\cdot\theta_1^3\cdot\theta_2^{-1},\\
\lambda_7=W\cdot\theta_1^2\cdot\theta_2^{-1},\\
\lambda_{8}=L\cdot\theta_1^{-1},\\
\lambda_{9}=R\cdot\theta_1^{-1}.
\end{cases}
\end{align}

The goal here is to find $\theta_1$ and $\theta_2$, as functions of $\pi_i$, $\forall i\in\{1,2,\ldots,7\}$, such that:
\begin{align}\label{eq:target_lambdas_3cst}
\begin{cases}
    \lambda_i\approx 1,\;\forall i\in U_3:=\{1,2,4,5,6,7,8,9\},\\
    \lambda_3<<1.
\end{cases}
\end{align}
That is, we would like to study the regime in which we can drop the $L_3$ term from the elastic energy.

We assume, as in \cite{rusconi25}, that the characteristic constants $\theta_1$ and $\theta_2$ can be written as a product of powers of the physical parameters $L_1$, $L_2$, $L_3$, $a$, $b$, $c$, $W$, $L$ and $R$. This assumption, together with \cite[Theorem 3]{rusconi25}, will guarantee that the coefficients for the rescaled problem $\lambda_1$, $\lambda_2$, $\ldots$, $\lambda_9$ can be written as power-law monomials of the governing parameters $\pi_1$, $\pi_2$, $\ldots$, $\pi_7$, that is:
\begin{align}\label{eq:lambda_pi}    \lambda_i=\pi_1^{\alpha_{i,1}}\cdot\pi_2^{\alpha_{i,2}}\cdot\pi_3^{\alpha_{i,3}}\cdot\pi_4^{\alpha_{i,4}}\cdot\pi_5^{\alpha_{i,5}}\cdot\pi_6^{\alpha_{i,6}}\cdot\pi_7^{\alpha_{i,7}},
\end{align}
for all $i\in\{1,2,\ldots,9\}$. 

\begin{figure}[htpb]
    \centering
\begin{tikzpicture}[x=0.125cm, y=-0.125cm]

\tikzset{
    val1/.style={fill=green, rectangle, inner sep=1.2pt}, 
    val2/.style={fill=blue, rectangle, inner sep=1.2pt},  
    val3/.style={fill=red, rectangle, inner sep=1.2pt}    
}

\draw[thick] (0.5,0.5) -- (0.5,49.5);
\draw[thick] (0.5,0.5) -- (64.5,0.5);
\draw[thick] (0.5,49.5) -- (64.5,49.5);
\draw[thick] (64.5,0.5) -- (64.5,49.5);

\foreach \x in {1, 10, 20, 30, 40, 50, 60, 64}
    \draw (\x, 49.5) -- (\x, 50) node[below] {\small \x};

\foreach \y in {1, 10, 20, 30, 40, 49}
    \draw (0, \y) -- (0.5, \y) node[left] {\small \y};

% S

% Green

\foreach \i in {1,...,14} \node[val1] at (\i, \i) {};

\foreach \i in {22,...,28} \node[val1] at (\i, \i) {};

\foreach \i in {29,...,49} \node[val1] at (\i+7, \i) {};

% Blue

\foreach \i in {1,...,14} \node[val2] at (\i+7, \i) {};

\foreach \i in {29,...,42} \node[val2] at (\i, \i) {};

\foreach \i in {22,...,28} \node[val2] at (\i+14, \i) {};

\foreach \i in {15,...,21} \node[val2] at (\i+28, \i) {};

\foreach \i in {36,...,49} \node[val2] at (\i+14, \i) {};

% Red

\foreach \i in {15,...,21} \node[val3] at (\i, \i) {};

\foreach \i in {15,...,21} \node[val3] at (\i+7, \i) {};

% s

\foreach \i in {7,13,17,25,33,37,39,43} \node[val2] at (64, \i) {};

\foreach \i in {14,32,38,44} \node[val1] at (64,\i) {};

\node[val3] at (64,20) {};

\draw (17, 55) rectangle (55, 60.5);
\node[val1] at (20, 57.75) {}; \node[right] at (22, 57.75) {\small -1};
\node[val2] at (32, 57.75) {}; \node[right] at (34, 57.75) {\small +1};
\node[val3] at (44, 57.75) {}; \node[right] at (46, 57.75) {\small -1/2};

\end{tikzpicture}
    \caption{The matrix $<\textbf{S}|\vec{s}>$ (white cells are $0$)}
    \label{fig:S|s_3cst}
\end{figure}
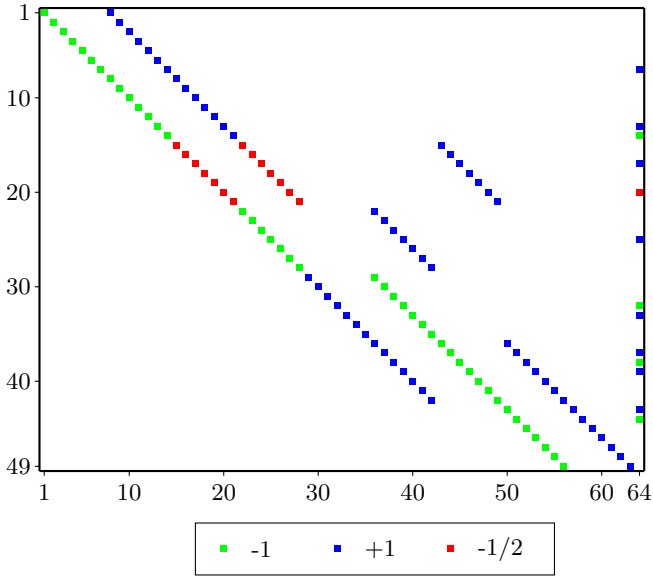

By \cite[Theorem 5]{rusconi25}, all the vectors $\vec{\alpha_i}$ must satisfy equation $\textbf{S}\vec{\alpha}=\vec{s}$, where both $\textbf{S}$ and $\vec{s}$ are computed explicitly based on our model. Since, in our case, $\textbf{S}\in\mathcal{M}_{49\times 63}(\mathbb{R})$ and $\vec{s}\in\mathcal{M}_{49\times 1}(\mathbb{R})$, we only draw the matrix $<\textbf{S}|\vec{s}>$ in \Cref{fig:S|s_3cst}, where the white cells are $0$.

In order to find the optimal values of $\vec{\alpha_i}\in\mathbb{R}^7$, for any $i\in\{1,2,\ldots,9\}$, we use the generalized optimal scaling technique from \cite{rusconi25}. That is, we consider $\vec{\alpha}$ as the minimizer of the following cost function, which is semipositive definite (hence, it admits a minimizer):
\begin{align}\label{eq:minimization_problem_GOS_3cst}
    C_{U_3}(\vec{\alpha})=\sum_{i\in U_3}\big(\log_{10}(\lambda_i)\big)^2,
\end{align}
where we recall that $U_3=\{1,2,\ldots,9\}\setminus\{3\}$. Here, $\lambda_i$ depends on $\pi_j$, based on \Cref{eq:lambda_pi}, hence, it depends on $\vec{\alpha}$. Moreover, the previous cost function reflects our desired model: we want all $\lambda_i\approx 1$ for $i\in U_3$ and we leave $\lambda_3$ out of the minimization problem by simply imposing, at the end, $\lambda_3<<1$. 

The existence of such optimal choice for the exponents $\vec{\alpha}$ is given by \cite[Theorem 7]{rusconi25}. To be more specific, an optimal choice for $\vec{\alpha}$ exists since there exists $\vec{\nu}$ such that the following system is satisfied:
\begin{align*}
    \begin{pmatrix}
        \Sigma & \textbf{0}_{49\times 49}\\
        \textbf{G} & \Sigma^T
    \end{pmatrix}\cdot\begin{pmatrix}
        \vec{\alpha}\\
        \vec{\nu}
    \end{pmatrix}=\begin{pmatrix}
        \vec{\sigma}\\
        \textbf{0}_{49\times 1}
    \end{pmatrix},
\end{align*}
where $\Sigma=\textbf{S}$, $\vec{\sigma}=\vec{s}$ and $\textbf{G}$ is the following matrix:
\begin{align*}
    \begin{pmatrix}
        \textbf{P} & \textbf{0}_{7\times 7} & \textbf{0}_{7\times 7} & \textbf{0}_{7\times 7} & \textbf{0}_{7\times 7} & \textbf{0}_{7\times 7} & \textbf{0}_{7\times 7} & \textbf{0}_{7\times 7} & \textbf{0}_{7\times 7}\\
        \textbf{0}_{7\times 7} & \textbf{P} & \textbf{0}_{7\times 7} & \textbf{0}_{7\times 7} & \textbf{0}_{7\times 7} & \textbf{0}_{7\times 7} & \textbf{0}_{7\times 7} & \textbf{0}_{7\times 7} & \textbf{0}_{7\times 7}\\
        \textbf{0}_{7\times 7} & \textbf{0}_{7\times 7} & \textbf{0}_{7\times 7} & \textbf{0}_{7\times 7} & \textbf{0}_{7\times 7} & \textbf{0}_{7\times 7} & \textbf{0}_{7\times 7} & \textbf{0}_{7\times 7} & \textbf{0}_{7\times 7}\\
        \textbf{0}_{7\times 7} & \textbf{0}_{7\times 7} & \textbf{0}_{7\times 7} & \textbf{P} & \textbf{0}_{7\times 7} & \textbf{0}_{7\times 7} & \textbf{0}_{7\times 7} & \textbf{0}_{7\times 7} & \textbf{0}_{7\times 7}\\
        \textbf{0}_{7\times 7} & \textbf{0}_{7\times 7} & \textbf{0}_{7\times 7} & \textbf{0}_{7\times 7} & \textbf{P} & \textbf{0}_{7\times 7} & \textbf{0}_{7\times 7} & \textbf{0}_{7\times 7} & \textbf{0}_{7\times 7}\\
        \textbf{0}_{7\times 7} & \textbf{0}_{7\times 7} & \textbf{0}_{7\times 7} & \textbf{0}_{7\times 7} & \textbf{0}_{7\times 7} & \textbf{P} & \textbf{0}_{7\times 7} & \textbf{0}_{7\times 7} & \textbf{0}_{7\times 7}\\
        \textbf{0}_{7\times 7} & \textbf{0}_{7\times 7} & \textbf{0}_{7\times 7} & \textbf{0}_{7\times 7} & \textbf{0}_{7\times 7} & \textbf{0}_{7\times 7} & \textbf{P} & \textbf{0}_{7\times 7} & \textbf{0}_{7\times 7}\\
        \textbf{0}_{7\times 7} & \textbf{0}_{7\times 7} & \textbf{0}_{7\times 7} & \textbf{0}_{7\times 7} & \textbf{0}_{7\times 7} & \textbf{0}_{7\times 7} & \textbf{0}_{7\times 7} & \textbf{P} & \textbf{0}_{7\times 7}\\
        \textbf{0}_{7\times 7} & \textbf{0}_{7\times 7} & \textbf{0}_{7\times 7} & \textbf{0}_{7\times 7} & \textbf{0}_{7\times 7} & \textbf{0}_{7\times 7} & \textbf{0}_{7\times 7} & \textbf{0}_{7\times 7} & \textbf{P}\\
    \end{pmatrix},
\end{align*}
where $\textbf{P}=\big(\textbf{p}_{ij}\big)_{i,j\in\overline{1,7}}$, with 
\begin{align*}
\textbf{p}_{ij}=2\log_{10}(\pi_i)\log_{10}(\pi_j),\; \forall i,j\in\overline{1,7}.
\end{align*}

We therefore obtain optimal exponents and, after replacing the governing parameters with the physical parameters, we get:
\begin{align*}
    \lambda_1&=\sqrt[41]{\dfrac{L_1^{26}}{L_2^{15}\cdot a\cdot b\cdot c\cdot L^7\cdot R^7\cdot W^8}}\approx 0.422\\
    \lambda_2&=\sqrt[41]{\dfrac{L_2^{26}}{L_1^{15}\cdot a\cdot b\cdot c\cdot L^7\cdot R^7\cdot W^8}}\approx 0.169\\
    \lambda_3&=\sqrt[41]{\dfrac{L_3^{41}}{L_1^{15}\cdot L_2^{15}\cdot a\cdot b\cdot c\cdot L^7\cdot R^7\cdot W^8}}\approx 0.338\\
    \lambda_4&=\sqrt[41]{\dfrac{a^{30}\cdot L^5\cdot R^5}{L_1\cdot L_2\cdot b^{11}\cdot c^{11}\cdot W^6}}\approx 0.699\\
    \lambda_5&=\sqrt[41]{\dfrac{b^{30}\cdot L^5\cdot R^5}{L_1\cdot L_2\cdot a^{11}\cdot c^{11}\cdot W^6}}\approx 8.662\\
    \lambda_6&=\sqrt[41]{\dfrac{c^{30}\cdot L^5\cdot R^5}{L_1\cdot L_2\cdot a^{11}\cdot b^{11}\cdot W^6}}\approx 7.036\\
    \lambda_7&=\sqrt[41]{\dfrac{W^{34}}{L_1^8\cdot L_2^8\cdot a^6\cdot b^6\cdot c^6\cdot L\cdot R}}\approx 0.331\\
    \lambda_8&=\sqrt[41]{\dfrac{L^{35}\cdot a^5\cdot b^5\cdot c^5}{L_1^7\cdot L_2^7\cdot R^6\cdot W}}\approx 48.841\\
    \lambda_9&=\sqrt[41]{\dfrac{R^{35}\cdot a^5\cdot b^5\cdot c^5}{L_1^7\cdot L_2^7\cdot L^6\cdot W}}\approx 12.210,
\end{align*}
where the approximations were computed using the val\-ues from \Cref{sec:initial_param}.

We first note that GOS is able to target the regime from \Cref{eq:target_lambdas_3cst}:
\begin{align*}
    10^{-2}<\lambda_i<10^2,\;\forall i\in\{1,2,4,5,6,7,8,9\}.
\end{align*}

Secondly, we can see, for example, that $L_3$, the elastic constant that we would like to eliminate from our model, appears only in $\lambda_3$, which was the only $\lambda$ not considered in the minimization problem from \Cref{eq:minimization_problem_GOS_3cst}.

Hence, $\lambda_3<<1$ implies the following inequality:
\begin{align*}
L_3<<\sqrt[41]{L_1^{15}\cdot L_2^{15}\cdot a\cdot b\cdot c\cdot L^7\cdot R^7\cdot W^8},
\end{align*}
which is exactly \Cref{eq:threshold_3cst}.

\begin{remark}
    If one is interested in studying other scale separation regimes for this configuration, then it would apply the same procedure as before for $U_{k}=\{1,2,3,4,5,6,7,8,9\}\setminus\{k\}$, where $k$ is the number of the component of $\vec{p}$ that we want to isolate. The critical values for each of these cases is highlighted in \Cref{tab:regimes_for_physical_param_3_cst}.
\end{remark}

\begin{table}[htbp]
\centering
\caption{Regimes for each physical parameters achieved using the GOS method}
\label{tab:regimes_for_physical_param_3_cst}
\begin{ruledtabular}
\begin{tabular}{cc}
        \textbf{Target} & \textbf{Regime}\\
        \hline
        $L_1$ & $L_1^{41}<<L_2^{15}\cdot L_3^{15}\cdot a\cdot b\cdot c\cdot L^7\cdot R^7\cdot W^8$\\
        \hline
        $L_2$ & $L_2^{41}<<L_1^{15}\cdot L_3^{15}\cdot a\cdot b\cdot c\cdot L^7\cdot R^7\cdot W^8$\\
        \hline
        $L_3$ & $L_3^{41}<<L_2^{15}\cdot L_3^{15}\cdot a\cdot b\cdot c\cdot L^7\cdot R^7\cdot W^8$\\
        \hline
        $a$ & $a^{41}<<L_1\cdot L_2\cdot L_3\cdot b^{15}\cdot c^{15}\cdot W^8\cdot L^{-7}\cdot R^{-7}$\\
        \hline
        $b$ & $b^{41}<<L_1\cdot L_2\cdot L_3\cdot a^{15}\cdot c^{15}\cdot W^8\cdot L^{-7}\cdot R^{-7}$\\
        \hline
        $c$ & $c^{41}<<L_1\cdot L_2\cdot L_3\cdot a^{15}\cdot b^{15}\cdot W^8\cdot L^{-7}\cdot R^{-7}$\\
        \hline
        $W$ & $W^6<<L_1\cdot L_2\cdot L_3\cdot a\cdot b\cdot c$\\
        \hline
        $L$ & $L^7<<L_1\cdot L_2\cdot L_3\cdot a^{-1}\cdot b^{-1}\cdot c^{-1}\cdot R$\\
        \hline
        $R$ & $R^7<<L_1\cdot L_2\cdot L_3\cdot a^{-1}\cdot b^{-1}\cdot c^{-1}\cdot L$
    \end{tabular}
    \end{ruledtabular}
\end{table}

\subsection{The $(L_1,L_2,0)\leadsto(L_1,0,0)$ case}

We start with the observation that $L_3$ is excluded from this case, since $L_3=0$. Hence, $\vec{p}$ is now:
\begin{align*}
\vec{p}:=(L_1,L_2,a,b,c,W,L,R)^T\in\mathbb{R}^{N_p},\;N_p=8.
\end{align*}

The matrix $\textbf{M}$ becomes:
\begin{align*}
    \textbf{M}=\begin{pmatrix}
        1 & 1 & 1 & 1 & 1 & 1 & 0 & 0\\
        1 & 1 & -1 & -1 & -1 & 0 & 1 & 1\\
        -2 & -2 & -2 & -2 & -2 & -2 & 0 & 0
    \end{pmatrix}
\end{align*}
and we have $\dim(\ker(\textbf{M}))=6:=N_0$. A basis is given by the column-vectors of the following matrix:
\begin{align*}
    \textbf{Z}=\begin{pmatrix}
        -\dfrac{1}{2} & -\dfrac{1}{2} & -\dfrac{1}{2} & 0 & 0 & -1\\
        0 & 0 & 0 & 0 & 0 & 1\\
        \dfrac{1}{2} & \dfrac{1}{2} & -\dfrac{1}{2} & -1 & -1 & 0\\
        0 & 0 & 0 & 0 & 1 & 0\\
        0 & 0 & 0 & 1 & 0 & 0\\
        0 & 0 & 1 & 0 & 0 & 0\\
        0 & 1 & 0 & 0 & 0 & 0\\
        1 & 0 & 0 & 0 & 0 & 0
    \end{pmatrix}.
\end{align*}

We have therefore only 6 governing parameters:
\begin{align*}
    &\pi_1=\dfrac{R\sqrt{a}}{\sqrt{L_1}},\;\pi_2=\dfrac{L\sqrt{a}}{\sqrt{L_1}},\;\pi_3=\dfrac{W}{\sqrt{L_1\cdot a}},\\
    &\hspace{10mm}\pi_4=\dfrac{c}{a},\;\pi_5=\dfrac{b}{a},\;\pi_6=\dfrac{L_2}{L_1}.
\end{align*}

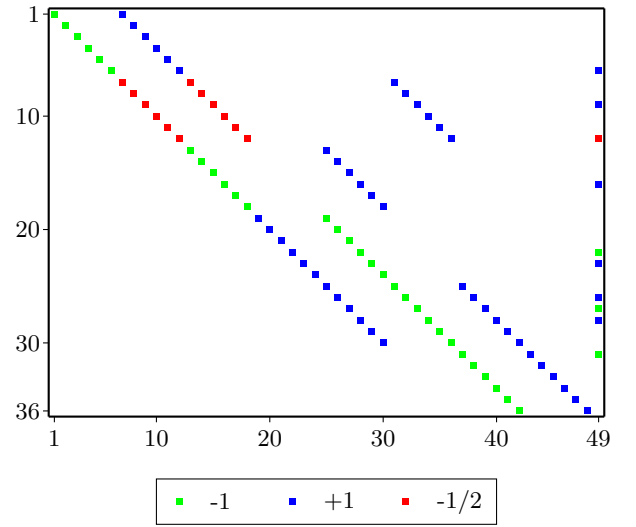
\begin{figure}[htpb]
    \centering
\begin{tikzpicture}[x=0.15cm, y=-0.15cm]

\tikzset{
    val1/.style={fill=green, rectangle, inner sep=1.2pt}, 
    val2/.style={fill=blue, rectangle, inner sep=1.2pt},  
    val3/.style={fill=red, rectangle, inner sep=1.2pt}   
}

\draw[thick] (0.5,0.5) -- (0.5,36.5);
\draw[thick] (0.5,0.5) -- (49.5,0.5);
\draw[thick] (0.5,36.5) -- (49.5,36.5);
\draw[thick] (49.5,0.5) -- (49.5,36.5);

\foreach \x in {1, 10, 20, 30, 40, 49}
    \draw (\x, 36.5) -- (\x, 37) node[below] {\small \x};

\foreach \y in {1, 10, 20, 30, 36}
    \draw (0, \y) -- (0.5, \y) node[left] {\small \y};

% Green
\foreach \i in {1,...,6} \node[val1] at (\i, \i) {};

\foreach \i in {13,...,18} \node[val1] at (\i, \i) {};

\foreach \i in {19,...,36} \node[val1] at (\i+6, \i) {};

% Blue

\foreach \i in {1,...,6} \node[val2] at (\i+6, \i) {};

\foreach \i in {19,...,30} \node[val2] at (\i, \i) {};

\foreach \i in {13,...,18} \node[val2] at (\i+12, \i) {};

\foreach \i in {7,...,12} \node[val2] at (\i+24, \i) {};

\foreach \i in {25,...,36} \node[val2] at (\i+12, \i) {};

% Red

\foreach \i in {7,...,12} \node[val3] at (\i, \i) {};

\foreach \i in {7,...,12} \node[val3] at (\i+6, \i) {};

% s

\foreach \i in {6,9,16,23,26,28,31} \node[val2] at (49, \i) {};

\foreach \i in {22,27,31} \node[val1] at (49,\i) {};

\node[val3] at (49,12) {};

\draw (10, 42) rectangle (40, 46);
 \node[val1] at (12, 44) {}; \node[right] at (14, 44) {\small -1};
 \node[val2] at (22, 44) {}; \node[right] at (24, 44) {\small +1};
 \node[val3] at (32, 44) {}; \node[right] at (34, 44) {\small -1/2};

\end{tikzpicture}
    \caption{The matrix $<\textbf{S}|\vec{s}>$ (white cells are $0$)}
    \label{fig:S|s_2cst}
\end{figure}

We rescale our problem in a similar fashion:
\begin{align*}
\overline{x}=\dfrac{x}{\theta_1}\;\text{and}\;\overline{\mathcal{F}}=\dfrac{\mathcal{F}}{\theta_2}
\end{align*}
and we have $8$ dimensionless coefficients:
\begin{align*}
\begin{cases}
\lambda_1=L_1\cdot\theta_1\cdot\theta_2^{-1},\\
\lambda_2=L_2\cdot\theta_1\cdot\theta_2^{-1},\\
\lambda_3=a\cdot\theta_1^3\cdot\theta_2^{-1},\\
\lambda_4=b\cdot\theta_1^3\cdot\theta_2^{-1},\\
\lambda_5=c\cdot\theta_1^3\cdot\theta_2^{-1},\\
\lambda_6=W\cdot\theta_1^2\cdot\theta_2^{-1},\\
\lambda_7=L\cdot\theta_1^{-1},\\
\lambda_8=R\cdot\theta_1^{-1}.
\end{cases}
\end{align*}
In a similar fashion, we are interested in obtaining the exponents $\alpha_{i,j}$, with $i\in\{1,2,\ldots,8\}$ and $j\in\{1,2,\ldots,6\}$, such that:
\begin{align*}
    \lambda_i=\pi_1^{\alpha_{i,1}}\cdot\pi_2^{\alpha_{i,2}}\cdot\pi_3^{\alpha_{i,3}}\cdot \pi_4^{\alpha{i,4}}\cdot\pi_5^{\alpha_{i,5}}\cdot\pi_6^{\alpha_{i,6}},
\end{align*}
for all $i\in\{1,2,\ldots,8\}$. In this case, the matrix $<\textbf{S}|\vec{s}>$ can be represented in \Cref{fig:S|s_2cst}, where once again the white cells are $0$.

We use a similar cost functional:
\begin{align}\label{eq:minimization_problem_GOS_2cst}
    C_{U_2}(\vec{\alpha})=\sum_{i\in U_2}\big(\log_{10}(\lambda_i)\big)^2,
\end{align}
where $U_2=\{1,2,\ldots,8\}\setminus\{2\}$. Here, we want:
\begin{align}\label{eq:target_lambdas_2cst}
\begin{cases}
    \lambda_i\approx 1,\;\forall i\in U_2:=\{1,3,4,5,6,7,8\},\\
    \lambda_2<<1.
\end{cases}
\end{align}

The consequent system becomes:
\begin{align*}
    \begin{pmatrix}
        \Sigma & \textbf{0}_{36\times 36}\\
        \textbf{G} & \Sigma^T
    \end{pmatrix}\cdot\begin{pmatrix}
        \vec{\alpha}\\
        \vec{\nu}
    \end{pmatrix}=\begin{pmatrix}
        \vec{\sigma}\\
        \textbf{0}_{36\times 1}
    \end{pmatrix},
\end{align*}
where $\Sigma=\textbf{S}$, $\vec{\sigma}=\vec{s}$ and $\textbf{G}$ is the following matrix:
\begin{align*}
    \begin{pmatrix}
        \textbf{P} & \textbf{0}_{6\times 6} & \textbf{0}_{6\times 6} & \textbf{0}_{6\times 6} & \textbf{0}_{6\times 6} & \textbf{0}_{6\times 6} & \textbf{0}_{6\times 6} & \textbf{0}_{6\times 6}\\
        \textbf{0}_{6\times 6} & \textbf{0}_{6\times 6} & \textbf{0}_{6\times 6} & \textbf{0}_{6\times 6} & \textbf{0}_{6\times 6} & \textbf{0}_{6\times 6} & \textbf{0}_{6\times 6} & \textbf{0}_{6\times 6}\\
        \textbf{0}_{6\times 6} & \textbf{0}_{6\times 6} & \textbf{P} & \textbf{0}_{6\times 6} & \textbf{0}_{6\times 6} & \textbf{0}_{6\times 6} & \textbf{0}_{6\times 6} & \textbf{0}_{6\times 6}\\
        \textbf{0}_{6\times 6} & \textbf{0}_{6\times 6} & \textbf{0}_{6\times 6} & \textbf{P} & \textbf{0}_{6\times 6} & \textbf{0}_{6\times 6} & \textbf{0}_{6\times 6} & \textbf{0}_{6\times 6}\\
        \textbf{0}_{6\times 6} & \textbf{0}_{6\times 6} & \textbf{0}_{6\times 6} & \textbf{0}_{6\times 6} & \textbf{P} & \textbf{0}_{6\times 6} & \textbf{0}_{6\times 6} & \textbf{0}_{6\times 6}\\
        \textbf{0}_{6\times 6} & \textbf{0}_{6\times 6} & \textbf{0}_{6\times 6} & \textbf{0}_{6\times 6} & \textbf{0}_{6\times 6} & \textbf{P} & \textbf{0}_{6\times 6} & \textbf{0}_{6\times 6}\\
        \textbf{0}_{6\times 6} & \textbf{0}_{6\times 6} & \textbf{0}_{6\times 6} & \textbf{0}_{6\times 6} & \textbf{0}_{6\times 6} & \textbf{0}_{6\times 6} & \textbf{P} & \textbf{0}_{6\times 6}\\
        \textbf{0}_{6\times 6} & \textbf{0}_{6\times 6} & \textbf{0}_{6\times 6} & \textbf{0}_{6\times 6} & \textbf{0}_{6\times 6} & \textbf{0}_{6\times 6} & \textbf{0}_{6\times 6} & \textbf{P}\\
    \end{pmatrix},
\end{align*}
where $\textbf{P}=\big(\textbf{p}_{ij}\big)_{i,j\in\overline{1,6}}$, with 
\begin{align*}
\textbf{p}_{ij}=2\log_{10}(\pi_i)\log_{10}(\pi_j),\; \forall i,j\in\overline{1,6}.
\end{align*}

The existence of the optimal exponents is once again ensured and, after replacing the governing parameters with the physical parameters, we get:
\begin{align*}
    \lambda_1&=\sqrt[26]{\dfrac{L_1^{11}}{a\cdot b\cdot c\cdot L^7\cdot R^7\cdot W^8}}\approx 0.204\\
    \lambda_2&=\sqrt[26]{\dfrac{L_2^{26}}{L_1^{15}\cdot a\cdot b\cdot c\cdot L^7\cdot R^7\cdot W^8}}\approx 0.060\\
    \lambda_3&=\sqrt[26]{\dfrac{a^{19}\cdot L^3\cdot R^3}{L_1\cdot b^7\cdot c^7\cdot W^4}}\approx 0.636\\
    \lambda_4&=\sqrt[26]{\dfrac{b^{19}\cdot L^3\cdot R^3}{L_1\cdot a^7\cdot c^7\cdot W^4}}\approx 7.834\\
    \lambda_5&=\sqrt[26]{\dfrac{c^{19}\cdot L^3\cdot R^3}{L_1\cdot a^7\cdot b^7\cdot W^4}}\approx 6.392\\
    \lambda_6&=\sqrt[26]{\dfrac{W^{20}}{L_1^8\cdot a^4\cdot b^4\cdot c^4\cdot L^2\cdot R^2}}\approx 0.044\\
    \lambda_7&=\sqrt[26]{\dfrac{L^{21}\cdot a^3\cdot b^3\cdot c^3}{L_1^7\cdot R^5\cdot W^2}}\approx 24.951\\
    \lambda_8&=\sqrt[26]{\dfrac{R^{21}\cdot a^3\cdot b^3\cdot c^3}{L_1^7\cdot L^5\cdot W^2}}\approx 6.238,
\end{align*}
which, once again, shows us that GOS is able to target the regime of lambdas described in \Cref{eq:target_lambdas_2cst}:
\begin{align*}
    10^{-2}<\lambda_i< 10^2,\;\forall i\in\{1,2,\ldots,8\}.
\end{align*}

We can see, for example, that $L_2$, the elastic constant that we want to eliminate from our model, appears only in $\lambda_2$, which was the only $\lambda$ not considered in the minimization problem from \Cref{eq:minimization_problem_GOS_2cst}. Hence, $\lambda_2<<1$ implies, in this case, the following inequality:
\begin{align*}
L_2<<\sqrt[26]{L_1^{15}\cdot a\cdot b\cdot c\cdot L^7\cdot R^7\cdot W^8}.
\end{align*}

\begin{remark}
    If one is interested in studying other scale separation regimes for this case, then the inequalities $\lambda_k<<1$, where $k$ is the position in $\vec{p}$ of the designated physical parameter, can be found in \Cref{tab:regimes_for_physical_param_2_cst}.
\end{remark}

\begin{table}[htbp]
\centering
    \caption{Regimes for each physical parameters achieved using the GOS method}
    \label{tab:regimes_for_physical_param_2_cst}
\begin{ruledtabular}
\begin{tabular}{cc}
        \textbf{Target} & \textbf{Regime}\\
        \hline
        $L_1$ & $L_1^{26}<<L_2^{15}\cdot a\cdot b\cdot c\cdot L^7\cdot R^7\cdot W^8$\\
        \hline
        $L_2$ & $L_2^{26}<<L_1^{15}\cdot a\cdot b\cdot c\cdot L^7\cdot R^7\cdot W^8$\\
        \hline
        $a$ & $a^{30}<<L_1\cdot L_2\cdot b^{11}\cdot c^{11}\cdot W^6\cdot L^{-5}\cdot R^{-5}$\\
        \hline
        $b$ & $b^{30}<<L_1\cdot L_2\cdot a^{11}\cdot c^{11}\cdot W^6\cdot L^{-5}\cdot R^{-5}$\\
        \hline
        $c$ & $c^{30}<<L_1\cdot L_2\cdot a^{11}\cdot b^{11}\cdot W^6\cdot L^{-5}\cdot R^{-5}$\\
        \hline
        $W$ & $W^{34}<<L_1^8\cdot L_2^8\cdot a^6\cdot b^6\cdot c^6\cdot L\cdot R$\\
        \hline
        $L$ & $L^{35}<<L_1^7\cdot L_2^7\cdot a^{-5}\cdot b^{-5}\cdot c^{-5}\cdot W\cdot R^6$\\
        \hline
        $R$ & $R^{35}<<L_1^7\cdot L_2^7\cdot a^{-5}\cdot b^{-5}\cdot c^{-5}\cdot W\cdot L^6$
    \end{tabular}
\end{ruledtabular}
\end{table}

\begin{table*}[htpb]
\caption{Numerical input data for the $(L_1, L_2, L_3) \leadsto (L_1, L_2, 0)$ case}
\label{tab:numerical_data_3cst}
\begin{ruledtabular}
\begin{tabular}{cccc}
$\lambda_3$ & $\texttt{L}_3$ & $\texttt{K}_1=\texttt{K}_3$ & $\texttt{K}_{24}$ \\
\colrule
$0$ & $0$ & $3.87596899224807$ & $3.22997416020672$\\
$10^{-7}$ & $0.00000059928435418997$ & $3.87596937511620$ & $3.22997454307485$\\
$10^{-6}$ & $0.00000599284354189969$ & $3.87597282092941$ & $3.22997798888806$\\
$10^{-5}$ & $0.00005992843541899690$ & $3.87600727906149$ & $3.23001244702014$\\
$10^{-4}$ & $0.00059928435418996900$ & $3.87635186038231$ & $3.23035702834096$\\
$10^{-3}$ & $0.00599284354189969000$ & $3.87979767359047$ & $3.23380284154912$\\
$10^{-2}$ & $0.05992843541899690000$ & $3.91425580567204$ & $3.26826097363069$\\
$10^{-1}$ & $0.59928435418996900000$ & $4.25883712648781$ & $3.61284229444646$\\
$1$ & $5.99284354189969000000$ & $7.70465033464546$ & $7.05865550260410$\\
\end{tabular}
\end{ruledtabular}
\end{table*}

\section{Numerical input data}\label{sec:numerical_data}

\subsection{The $(L_1,L_2,L_3)\to(L_1,L_2,0)$ case}

Here, we started with $K_{24}=7$ pN, so, after rescalings, we have
\begin{align*}
    \texttt{K}_{24}\approx 4.5219638242894.
\end{align*}
Using \Cref{eq:el_cst_OF=LdG}, we obtain:
\begin{align}\label{eq:openQmin_3cst_fixed_L1_L2}
\begin{cases}
    \texttt{L}_1\approx 2.52785855695909,\\
    \texttt{L}_2\approx 1.01114342278365,\\
    \texttt{L}_3\approx 2.02228684556726.
\end{cases}
\end{align}

In this dimensionless rescaled setting, the critical value from \Cref{eq:threshold_3cst} has the same form as before, but now with rescaled factors, due to \Cref{sec:invariance}:
\begin{align}\label{eq:threshold_3cst_openQmin}
    \texttt{L}_3<<\sqrt[41]{\texttt{L}_1^{15}\cdot \texttt{L}_2^{15}\cdot \texttt{a}\cdot \texttt{b}\cdot \texttt{c}\cdot \texttt{W}^8\cdot \texttt{L}^7\cdot \texttt{R}^7}:=\overline{\texttt{L}_3}
\end{align}
and we get
\begin{align*}
    \overline{\texttt{L}_3}\approx 5.99284354189972.
\end{align*}

Since \texttt{open-Qmin} has as input the Oseen-Frank constants, we make use of \Cref{eq:el_cst_OF=LdG} to determine all the numerical values for the \texttt{K}'s, depending on our choice of $\lambda_3$, which can be found in \Cref{tab:numerical_data_3cst}. Since $\texttt{L}_1$ and $\texttt{L}_2$ remain fixed as in \Cref{eq:openQmin_3cst_fixed_L1_L2}, then, whenever we change $\texttt{L}_3$, we only modify $\texttt{K}_1(=\texttt{K}_3)$ and $\texttt{K}_{24}$, so $\texttt{K}_{2}$ also stays fixed, that is:
\begin{align*}
    \texttt{K}_2\approx 3.22997416020672.
\end{align*}

\subsection{The $(L_1,L_2,L_3)\to(L_1,L_2,0)$ case}

Here, we started with $K_{24}=K_2=5$ pN. Using \Cref{eq:el_cst_OF=LdG}, we obtain:
\begin{align}\label{eq:openQmin_2cst_fixed_L1_L3}
\begin{cases}
    \texttt{L}_1\approx 2.52785855695909,\\
    \texttt{L}_2\approx 3.03343026835091,\\
    \texttt{L}_3=0.
\end{cases}
\end{align}

In this dimensionless rescaled setting, the critical value from \Cref{eq:threshold_2cst} has the same form as before, but now with rescaled factors, due to a similar argument as in \Cref{sec:invariance}:
\begin{align}\label{eq:threshold_2cst_openQmin}
    \texttt{L}_2<<\sqrt[26]{\texttt{L}_1^{15}\cdot \texttt{a}\cdot \texttt{b}\cdot \texttt{c}\cdot \texttt{L}^7\cdot \texttt{R}^7\cdot \texttt{W}^8}:=\overline{\texttt{L}_2}
\end{align}
and we obtain:
\begin{align*}
    \overline{\texttt{L}_2}\approx 16.7297994428135.
\end{align*}

We recall here that $\texttt{L}_1$ and $\texttt{L}_3$ remain fixed as in \Cref{eq:openQmin_2cst_fixed_L1_L3} and we only modify $\texttt{L}_2$. This implies that we are changing only the values of $\texttt{K}_1=\texttt{K}_3$, while $\texttt{K}_2=\texttt{K}_{24}$ remain fixed. All the numerical input data for our choices of $\lambda_2$ and the corresponding elastic constants can be found in \Cref{tab:numerical_data_2cst}.

\begin{remark}
All data from \Cref{tab:numerical_data_3cst,tab:numerical_data_2cst} can also be found in the supplementary material in the Zenodo repository \cite{zenodo}.
\end{remark}

\begin{table}[t]
\caption{Numerical input data for the $(L_1, L_2, 0) \leadsto (L_1, 0, 0)$ case}
\label{tab:numerical_data_2cst}
\begin{ruledtabular}
\begin{tabular}{ccc}
$\lambda_2$ & $\texttt{L}_2$ & $\texttt{K}_1=\texttt{K}_3$ \\
\colrule
$0$ & $0$ & $3.22997416020672$ \\
$10^{-7}$ & $0.00000167297994428135$ & $3.22997522903274$\\
$10^{-6}$ & $0.00001672979944281350$ & $3.2299848484669$\\
$10^{-5}$ & $0.00016729799442813500$ & $3.23008104280851$\\
$10^{-4}$ & $0.00167297994428135000$ & $3.23104298622464$\\
$10^{-3}$ & $0.01672979944281350000$ & $3.24066242038587$\\
$10^{-2}$ & $0.16729799442813500000$ & $3.3368567619982$\\
$10^{-1}$ & $1.67297994428135000000$ & $4.29880017812151$\\
$1$ & $16.72979944281350000000$ & $13.9182343393546$\\
\end{tabular}
\end{ruledtabular}
\end{table}

\section{Invariance of critical thresholds with respect to rescalings}\label{sec:invariance}

Let us suppose we have the inequality from \Cref{eq:threshold_3cst}, that is:
\begin{align*}
    L_3<<\sqrt[41]{L_1^{15}\cdot L_2^{15}\cdot a\cdot b\cdot c\cdot W^8\cdot L^7\cdot R^7}:=\overline{L_3}
\end{align*}
and the rescalings of the type \Cref{eq:change_of_variables} for the \texttt{open-Qmin} values, that is:
\begin{align*}
\begin{cases}
\texttt{L}_1=L_1\cdot\theta_1\cdot\theta_2^{-1},\\
\texttt{L}_2=L_2\cdot\theta_1\cdot\theta_2^{-1},\\
\texttt{L}_3=L_3\cdot\theta_1\cdot\theta_2^{-1},\\
\texttt{a}=a\cdot\theta_1^3\cdot\theta_2^{-1},\\
\texttt{b}=b\cdot\theta_1^3\cdot\theta_2^{-1},\\
\texttt{c}=c\cdot\theta_1^3\cdot\theta_2^{-1},\\
\texttt{W}=W\cdot\theta_1^2\cdot\theta_2^{-1},\\
\texttt{L}=L\cdot\theta_1^{-1},\\
\texttt{R}=R\cdot\theta_1^{-1}.
\end{cases}
\end{align*}

Then the critical threshold can be written as:
\begin{align*}
    &\overline{L_3}^{41}=L_1^{15}\cdot L_2^{15}\cdot a\cdot b\cdot c\cdot W^8\cdot L^7\cdot R^7\\
    &=\big(\texttt{L}_1\cdot\theta_1^{-1}\cdot\theta_2\big)^{15}\cdot\big(\texttt{L}_1\cdot\theta_1^{-1}\cdot\theta_2\big)^{15}\cdot\\
    &\;\;\;\;\;\cdot\texttt{a}\cdot\theta_1^{-3}\cdot\theta_2\cdot\texttt{b}\cdot\theta_1^{-3}\cdot\theta_2\cdot\texttt{c}\cdot\theta_1^{-3}\cdot\theta_2\cdot\\
    &\;\;\;\;\; \cdot\big(\texttt{W}\cdot\theta_1^{-2}\cdot\theta_2\big)^8\cdot\big(\texttt{L}\cdot\theta_1\big)^7\cdot\big(\texttt{R}\cdot\theta_1\big)^7\\
    &=\big(\texttt{L}_1^{15}\cdot\texttt{L}_2^{15}\cdot\texttt{a}\cdot\texttt{b}\cdot\texttt{c}\cdot\texttt{W}^8\cdot\texttt{L}^7\cdot\texttt{R}^7\big)\cdot\theta_1^{-41}\cdot\theta_2^{41}\\
    &=\overline{\texttt{L}_3}^{41}\cdot\theta_1^{-41}\cdot\theta_2^{41},
\end{align*}
which is equivalent to
\begin{align*}
    \overline{\texttt{L}_3}=\overline{L}_3\cdot\theta_1\cdot\theta_2^{-1}.
\end{align*}

The last equation represents the rescaling of the critical dimensionless threshold $\overline{L}_3$ into the one for the \texttt{open-Qmin} values, which is $\overline{\texttt{L}_3}$, hence, \Cref{eq:threshold_3cst} becomes indeed \Cref{eq:threshold_3cst_openQmin}.

In a similar fashion, all other thresholds found by GOS can be rescaled and computed for the values used for \texttt{open-Qmin} simulations.

\end{appendix}

\bibliography{apssamp}

\end{document}